\documentclass[preprint]{aastex}
\usepackage{emulateapj5}
\usepackage[]{natbib}
 
\newcommand{\etal}{et al.}
\newcommand{\per}{\ensuremath{^{-1}}}
\newcommand{\persq}{\ensuremath{^{-2}}}

\newcommand{\percucm}{cm\ensuremath{^{-3}}}
\newcommand{\hal}{H\ensuremath{\alpha}}

\newcommand{\hst}{\emph{HST}}
\newcommand{\msun}{\ensuremath{M_{\odot}}}
\newcommand{\kms}{km s\ensuremath{^{-1}}}

\newcommand{\ncrit}{\ensuremath{n_\mathrm{crit}}}
\newcommand{\lam}{\ensuremath{\lambda}}

\newcommand{\lya}{Ly \ensuremath{\alpha}}
\newcommand{\hn}{\hal+[\ion{N}{2}]}
\newcommand{\alphaox}{\ensuremath{\alpha_\mathrm{ox}}}
\newcommand{\rin}{\ensuremath{r_\mathrm{in}}}
\newcommand{\rs}{\ensuremath{R_\mathrm{s}}}
\newcommand{\mbh}{\ensuremath{M_\mathrm{BH}}}
 
\slugcomment{To appear in The Astrophysical Journal}
\shorttitle{BLR AND NLR OF NGC 4579}
\shortauthors{BARTH ET AL.}
 
\begin{document}
 
\title{The Broad-Line and Narrow-Line Regions of the LINER NGC
4579\footnotemark[1]}
 
\footnotetext[1]{Based on observations with the NASA/ESA \emph{Hubble
Space Telescope} obtained at STScI, which is operated by AURA, Inc.,
under NASA contract NAS5-26555.}

\author{Aaron J. Barth\altaffilmark{2}, Luis C. Ho\altaffilmark{3},
Alexei V. Filippenko\altaffilmark{4}, Hans-Walter Rix\altaffilmark{5}, \\
and Wallace L. W. Sargent\altaffilmark{6}}
 
\altaffiltext{2}{Harvard-Smithsonian Center for Astrophysics, 60 Garden Street,
Cambridge, MA 02138.}

\altaffiltext{3}{The Observatories of the Carnegie Institution of Washington,
813 Santa Barbara Street, Pasadena, CA 91101.}

\altaffiltext{4}{Department of Astronomy, University of California,
Berkeley, CA 94720-3411.}
 
\altaffiltext{5}{Max-Planck-Institut f\"ur Astronomie, K\"onigstuhl 17,
Heidelberg D-69117, Germany.}
 
\altaffiltext{6}{Palomar Observatory, Caltech 105-24, Pasadena, CA 91125.}

\begin{abstract}
We report the discovery of an extremely broad \hal\ emission line in
the LINER nucleus of NGC 4579.  From ground-based observations, the
galaxy was previously known to contain a Type 1 nucleus with a broad
\hal\ line of FWHM $\approx 2300$ \kms\ and FWZI $\approx 5000$ \kms.
New spectra obtained with the \emph{Hubble Space Telescope} (\hst) and
a 0\farcs2-wide slit reveal an \hal\ component with a FWZI of $\sim
18,\!000$ \kms.  The line is not obviously double-peaked, but it does
possess ``shoulders'' on the red and blue sides which resemble the
\hal\ profiles of double-peaked emitters such as NGC 4203 and NGC
4450.  This similarity suggests that the very broad \hal\ profile in
NGC 4579 may represent emission from an accretion disk.  Three such
objects have been found recently in two \hst\ programs which have
targeted a total of 30 galaxies, demonstrating that double-peaked or
extremely broad-line emission in LINERs must be much more common than
would be inferred from ground-based surveys.  The ratio of the narrow
[\ion{S}{2}] \lam\lam6716, 6731 lines shows a pronounced gradient
indicating a steep rise in density toward the nucleus.  The direct
detection of a density gradient within the inner arcsecond of the
narrow-line region confirms expectations from previous observations of
linewidth-critical density correlations in several LINERs.

\end{abstract}


\section{Introduction}

In the twenty years since LINERs were first identified as a major
component of the extragalactic population \citep{h80}, most research
on these objects has been directed toward resolving the question of
whether they represent a type of accretion-powered active galactic
nucleus (AGN), or whether they are instead powered by purely stellar
phenomena.  The LINER class may be heterogeneous, but it is now very
clear that at least some LINERs must be genuine AGNs \citep[for a
review, see][]{ho99a}.  An exciting recent development has been the
detection of very broad, double-peaked \hal\ emission lines in a
handful of nearby LINERs \citep{sbw93, bow96, shi00, ho00}, a
phenomenon which is most readily interpreted in terms of emission from
a relativistic accretion disk.

NGC 4579 is a Virgo cluster Sb-type spiral with a redshift of $cz =
1520$ \kms\ and a distance of 21 Mpc according to recent Tully-Fisher
measurements \citep{gav99}.  It has long been known to host a LINER
nucleus with broad wings on the \hal\ emission-line profile
\citep{sta82, kee83, fs85}.  Using a ground-based spectrum obtained at
the Palomar 5-m telescope, \citet{hfs97b} found that the broad
component of \hal\ has a full-width at half maximum (FWHM) of $2300$
\kms.  \emph{Hubble Space Telescope} (\hst) Faint Object Spectrograph
data have revealed substantially broader permitted lines in the
ultraviolet (UV): for example, \ion{C}{4} \lam1549 has FWHM
$=6600$ \kms\ and full-width at zero intensity (FWZI) of
$13,\!000$ \kms\ \citep{bar96}.  In this paper, we describe an \hst\
optical spectrum of the NGC 4579 nucleus, which provides new insights
into the properties of this low-luminosity AGN.  Our analysis of the
rather complex emission-line kinematics will be presented elsewhere.

\section{Observations}

We observed NGC 4579 with the Space Telescope Imaging Spectrograph
(STIS) on 1999 April 21 UT as part of a program to study the
emission-line kinematics of a sample of six nearby AGNs.  The galaxy
was observed at five parallel positions with the STIS 0\farcs2-wide
slit, with a gap of 0\farcs05 between adjacent slit positions.  Six
spacecraft orbits were divided roughly equally among the five slit
positions, and the exposure time for the central slit position was
2700 s.  The telescope was oriented with the STIS slit at a position
angle of 95\fdg3.

The observations were obtained with the G750M grating and a tilt
setting of 6300--6860 \AA, giving coverage of [\ion{O}{1}]
\lam\lam6300, 6363, the \hn\ emission blend, and the [\ion{S}{2}]
\lam\lam6716, 6731 doublet.  For this setting, the point-source
spectral resolution is $\sim0.9$ \AA.  The CCD was read out in
unbinned mode, for a pixel scale of 0.554 \AA\ along the dispersion
direction and 0\farcs0507 along the spatial axis.  Three separate
exposures were combined with rejection of cosmic-ray hits.  The data
were calibrated by the standard pipeline at STScI.  To remove
remaining hot pixels, we performed one additional step of cleaning the
data using a filtering algorithm which replaced deviant pixels with a
local median value.  The cleaned images were then geometrically
rectified using the STIS calibration software.

\section{Broad \hal\ Emission}

\subsection{Observed Properties}

\begin{figure}
\plotone{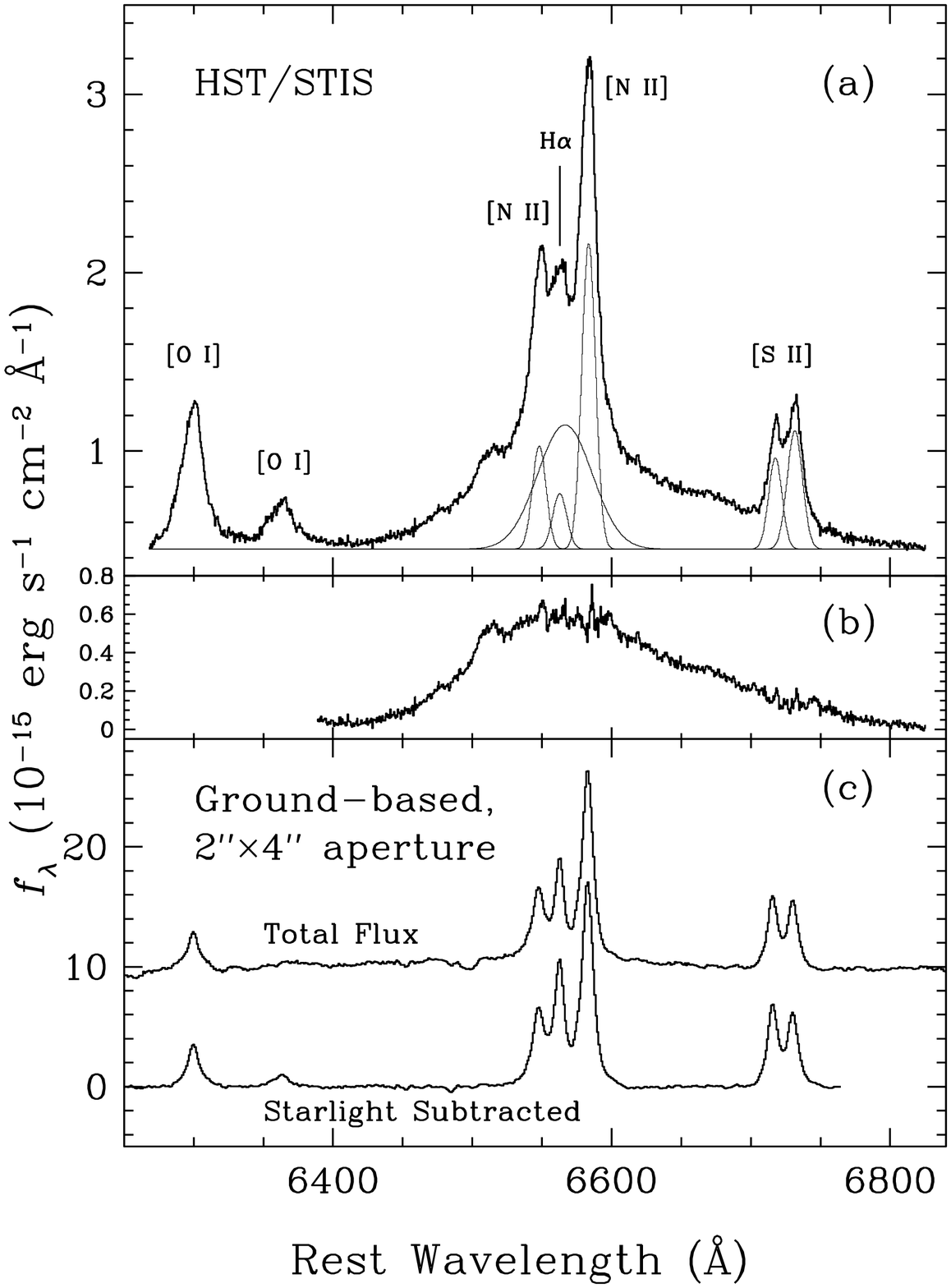}
\caption{Spectra of the nucleus of NGC 4579.  (\emph{a}) STIS G750M
spectrum through a $0\farcs2 \times 0\farcs2$ aperture.  The Gaussian
fits to the narrow lines and the ``normal'' broad \hal\ component are
shown.  (\emph{b}) The \hal\ profile after subtraction of the Gaussian
narrow and broad components.  (\emph{c}) Ground-based spectra from
\citet{hfs97a}, before and after starlight subtraction.
\label{fig1}}
\end{figure}

Figure \ref{fig1} shows the \hst\ spectrum of the NGC 4579 nucleus.
The extraction is centered on the nucleus and is a sum of four CCD
rows (0\farcs2 total) in the 2-dimensional spectrum.  The background
level (including starlight from the galaxy bulge) is negligible in
comparison with the nuclear brightness, and no background subtraction
has been applied.  In addition to the usual narrow-line features
expected in a LINER spectrum, there is a surprisingly broad and
asymmetric component to the \hal\ emission.  On the blue side, the
profile has a ``shoulder'' at 6520 \AA, just blueward of the base of
the [\ion{N}{2}] \lam6548 line.  The red side of the profile does not
exhibit such well-defined features, although there is a shallow
shoulder on the profile at $\sim6670$ \AA.  These shoulders are real
features and not merely artifacts of the data reduction or geometric
rectification, as they also appear in spectral extractions of the raw
data, even for large extraction widths.  The blue edge of the line
appears to be at $\sim6420$ \AA\ rest wavelength, while its red tail
extends to the end of the spectrum at 6820 \AA, giving FWZI $\approx
18,\!000$ \kms.  For comparison, Figure \ref{fig1} also displays the
ground-based spectrum from Ho, Filippenko, \& Sargent (1997a), which
was obtained at the Palomar 5-m telescope on 1984 February 12 UT.

We performed profile fits to remove the narrow-line contributions of
\hal, [\ion{N}{2}], and [\ion{S}{2}].  To fit the narrow lines, we
first fit a spline to the approximate shape of the very broad \hal\
emission profile, and subtracted this component from the spectrum,
leaving only the narrow lines and a possible contribution from a
``normal'' broad \hal\ component.  The [\ion{S}{2}] doublet was fitted
with a pair of Gaussians.  For the \hn\ blend, we used a four-Gaussian
fit representing narrow and broad \hal\ and the two [\ion{N}{2}]
lines, which were constrained to have the same velocity width and a
height ratio of 3:1.  The narrow \hal, [\ion{N}{2}], and [\ion{S}{2}]
lines have FWHM = 520--570 \kms, while the normal broad component of
\hal\ has FWHM = 2100 \kms, very similar to the value of 2300 \kms\
found by \citet{hfs97a}.

After subtracting the fitted components, the remaining very broad
component of \hal\ has $f = 1.1\times10^{-13}$ erg cm\persq\ s\per,
corresponding to a luminosity of $5.8\times10^{39}$ erg s\per\ for a
distance of 21 Mpc.  Since there have been no previous optical \hst\
spectra of the NGC 4579 nucleus, it is not possible to determine
whether this component has always been present or whether it is a
transient feature.  The amplitude of the broad feature is sufficiently
low that it could be present but undetected in the Palomar spectrum.
In general, it is extremely difficult to detect such broad,
low-amplitude features in ground-based spectra, because of the
overwhelming contamination by starlight from the bulge of the host
galaxy.  To make matters worse, elliptical galaxies and spiral bulges
generally display a broad continuum hump in the region 6400-6700 \AA\
\citep[e.g.,][]{fs85}.  The starlight subtraction method employed by
\citet{hfs97a} was designed to yield zero continuum flux in the
subtracted spectrum, and this procedure could have removed a faint,
very broad \hal\ feature by oversubtracting the stellar continuum
template.

\begin{figure}
\plotone{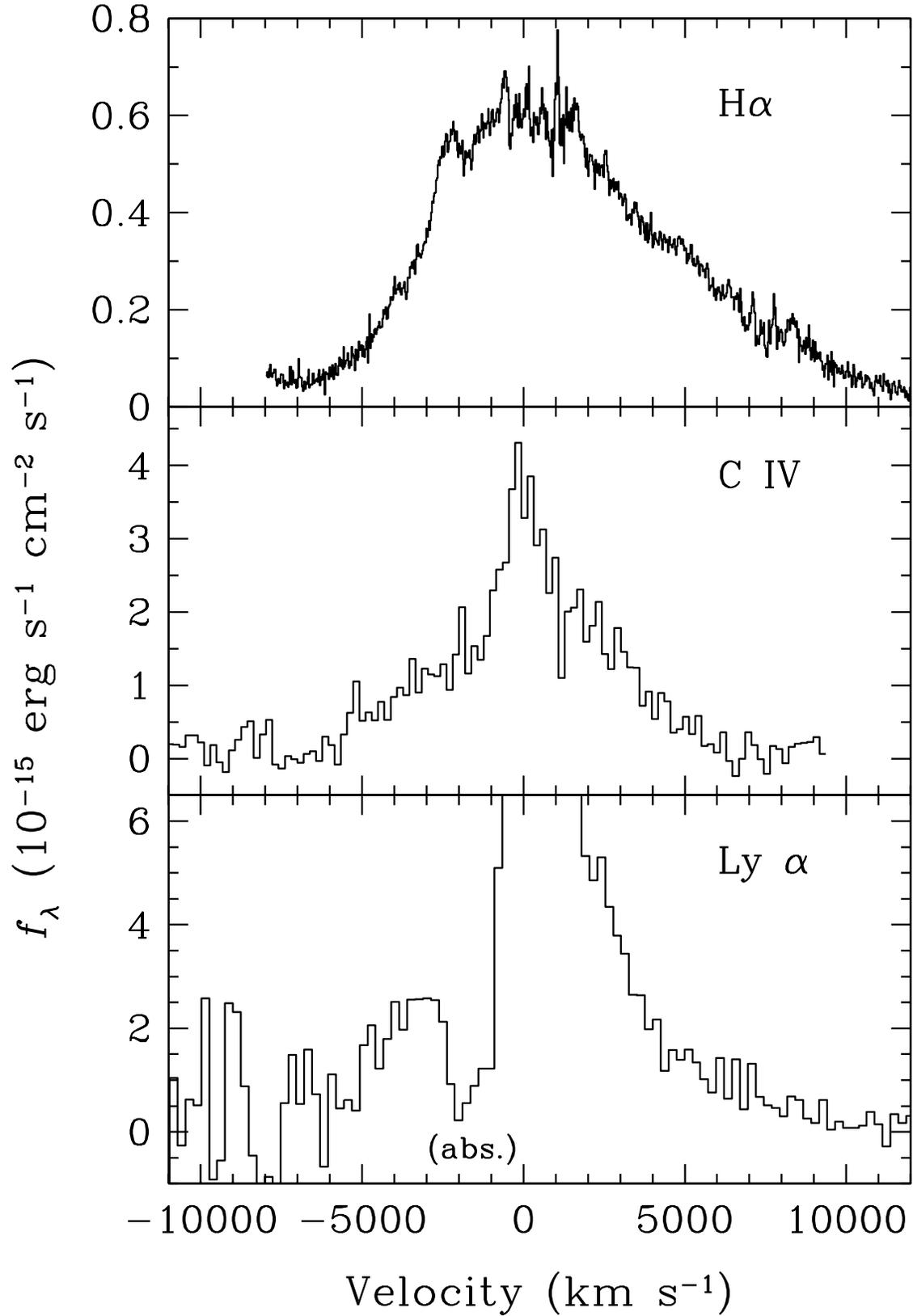}
\caption{Comparison of the broad-line profiles of \hal, C IV \lam1549,
and \lya. The blue side of the \lya\ profile is affected by Galactic
absorption, and a spike of geocoronal emission has been removed from
the spectrum.  The narrow components of \lya\ and C IV have not been
removed.
\label{fig2}}
\end{figure}

No other broad optical lines have been detected in NGC 4579, but the
\hst\ UV spectrum (taken in 1994 December) clearly demonstrates that
other permitted lines are very broad as well.  The UV spectrum has low
signal-to-noise ratio (S/N) in the grating setting that covers \lya\
and \ion{C}{4}, but emission from background starlight in this
wavelength range is negligible.  As shown in Figure \ref{fig2}, the
\lya\ and \ion{C}{4} profiles have blue wings which extend to
$\sim6000$ \kms\ from the systemic velocity, comparable to the maximum
extent of the blue wing of \hal.  The red wing of \hal\ extends to
velocities roughly 4000 \kms\ greater than the red wing of \ion{C}{4}.
It is difficult to discern the maximum extent of the red wing of \lya\
due to the low S/N and a possible contribution from \ion{N}{5}
\lam1240, but \lya\ emission is seen out to perhaps $9000$ \kms\
from the systemic velocity.

\subsection{Relationship to the Central Engine}

Double-peaked broad \hal\ emission lines have previously been detected
in six LINERs as well as in several radio galaxies and radio-loud
quasars; see \citet{eh94}, \citet{fil96}, and \citet{ho00} for reviews
of their properties.  Models of emission from rotating, Keplerian
accretion disks provide the best explanation for the origin of the
double-peaked \hal\ lines in these objects \citep[e.g.,][]{hf88,
chf89, hal96, sb97}.  The width of the broad-line profile in NGC 4579
is comparable to the linewidths in these double-peaked emitters, which
have typical FWZI $\approx 15,\!000-20,\!000$ \kms.  While the NGC
4579 \hal\ profile is not clearly double-peaked, the shoulders at
$-2500$ and 5000 \kms\ are very similar to the features detected in
NGC 4203 \citep{shi00} and NGC 4450 \citep{ho00}.  The overall
similarities in the line profiles suggest a common origin for the
broad-line emission in these three galaxies.  Their profile shoulders
are much less pronounced than the broad humps in ``classic''
double-peaked emitters such as Arp 102b \citep[e.g.,][]{hf88}, but
\citet{ho00} have shown that a relativistic accretion disk model can
fit the double-peaked profile in NGC 4450.

Assuming that the very broad emission from NGC 4579 originates in an
accretion disk, the inner radius of the line-emitting portion of the
disk can be estimated from the maximum velocity extent of the line.
For a disk inclination of 45\arcdeg, the projected velocity of the red
wing of \hal\ ($12,\!000$ \kms) corresponds to a radius of 160
Schwarzschild radii (\rs).  The very broad line does not uniquely
imply the presence of an accretion disk, however.  Since the line does
not possess a disklike, strongly double-peaked profile, other models
may be able to explain the \hal\ profile.  Spherical broad-line region
(BLR) models incorporating radial outflows can produce line profiles
with horns and shoulders as well as very broad wings \citep{rpb90},
qualitatively resembling the \hal\ profile in NGC 4579.  In addition,
the broad-line profiles produced by the outflow models possess an
intermediate-width core similar in appearance to the ``normal''
broad-line component observed in NGC 4579.  If the accretion disk
model does apply to NGC 4579, then the intermediate-width component of
\hal\ most likely originates in the ``normal'' BLR of NGC 4579 rather
than in the accretion disk, as the disk alone will produce a
double-horned rather than centrally peaked profile
\citep[e.g.,][]{erac95}.  Linewidths of $\sim2000$ \kms\ are typical
for the broad \hal\ components seen in ground-based spectra of LINERs
\citep{hfs97b}.

Alternatively, models based on binary black holes \citep{gas83} can
also produce very broad, asymmetric emission-line profiles.  Such
models have been ruled out, at least for a few well-studied objects,
because the velocities of the broad wings do not exhibit the pattern
of long-term variability that would be expected from orbital motion of
a close pair of black holes \citep{erac97}.

NGC 4579 shares other properties in common with the double-peaked
emitters, most notably the overall shape of its spectral energy
distribution (SED).  In particular, NGC 4579 lacks a ``big blue bump''
component, and its radio-to-optical flux ratio qualifies it as
radio-loud although its radio luminosity on an absolute scale is small
\citep{ho99b}.  Advection-dominated accretion flows
\citep[ADAFs;][]{ny94} have been proposed as an explanation for these
properties, and \citet{qua99} have shown that the SED of NGC 4579 can
be explained by emission from a thin accretion disk with a transition
at $\sim100\rs$ to an ADAF for the innermost accretion flow.  By
truncating the thin-disk portion of the accretion flow and removing
the innermost, hottest portion of the thin disk, the models are able
to produce an SED without a big blue bump.  Interior to the truncation
radius, the disk is presumed to evaporate into an ADAF, and this
portion of the accretion flow generates the X-ray emission. The broad
\hal\ linewidth is consistent with this scenario, as it requires the
truncation radius of the thin disk to be at $r \lesssim 160$ \rs.

However, the value of \mbh\ used by \citet{qua99} in modeling the SED
of NGC 4579 is probably too low by an order of magnitude.  The
black-hole mass can be estimated using the recently discovered
correlation between \mbh\ and bulge velocity dispersion \citep{fm00,
geb00}.  Central velocity dispersion measurements for NGC 4579 range
from 130 to 187 \kms\ \citep{wmt85, gdp96, pal97, hs98}.  Applying the
\citet{geb00} and \citet{fm00} relations, the predicted black-hole
mass is in the range $(1.8-9.3)\times10^7$ \msun.\footnote{In applying
the \citet{geb00} relation, we neglect the small ($\lesssim10\%$)
difference between the central velocity dispersion and the
luminosity-weighted velocity dispersion within the effective radius,
as spatially resolved measurements of $\sigma$ are not available.}
The smaller value of $4\times10^6$ \msun\ used by \citet{qua99} was a
crude estimate based on the UV broad-line widths and ionization
parameter \citep{bar96}, and the $\mbh-\sigma$ correlations probably
predict the central mass more accurately.  With this larger central
mass, the model predictions for the SED are altered, and it may be
possible to fit the UV/optical portion of the SED with a thin disk
extending all the way in to $\rin = 3\rs$, as described by
\citet{qua99} for the case of M81.  It remains to be seen whether
models of a truncated thin disk surrounding an ADAF can be constructed
to simultaneously fit the UV/optical continuum and the X-ray
luminosity of NGC 4579, for a central mass of $\mbh \approx
5\times10^7$ \msun.

Other evidence for an accretion disk in NGC 4579 is ambiguous.  Iron
K-line emission observed in \emph{ASCA} spectra \citep{ter98, ter00}
may originate from the inner accretion disk.  However, the line does
not clearly show a disk-like, relativistically broadened profile like
those seen in some Seyfert 1 nuclei \citep[e.g.,][]{tan95, nan99},
which would be expected if the disk extended all the way in to the
black hole.  Future X-ray observations with greater sensitivity and
higher spectral resolution can better constrain the nature of the
inner accretion flow and the possible presence of an ADAF.

\section{Narrow-line Properties}

Shock heating was the first mechanism proposed for the excitation of
LINER narrow-line regions (NLRs), largely because observations of the
[\ion{O}{3}] \lam\lam4959, 5007 and \lam4363 lines seemed to indicate
electron temperatures higher than would be expected for photoionized
gas \citep[e.g.,][]{ko76, fos78}.  Subsequent studies demonstrated
that the observed [\ion{O}{3}] line ratios could be reproduced by
photoionization models provided that a high-density component ($n_e
\gtrsim 10^{5.5}$ \percucm) is present in the NLR to boost emission in
the [\ion{O}{3}] \lam4363 line \citep{fh84}.  Correlations between
narrow-line width and critical density for collisional deexcitation
(\ncrit), found in several LINERs, support this hypothesis by
demonstrating the probable presence of radial density gradients in the
NLR \citep{fh84, fil85, fs88}.  Due to the difficulty of obtaining
spectra with subarcsecond spatial resolution, however, there has been
little \emph{direct} evidence (in the form of spatially resolved
density measurements) for density stratification in LINERs.  The STIS
observations present an opportunity to revisit this issue, using the
ratio of the [\ion{S}{2}] \lam\lam6716, 6731 lines as a density
diagnostic.

Since the S/N of the observations is not sufficient for fitting the
[\ion{S}{2}] lines on a row-by-row basis, we performed multi-row
extractions at several locations along the slit, for all five slit
positions.  When possible, we extracted regions corresponding to a
well-defined cloud or clump.  For each extraction region, we
subtracted the local background level and fitted the [\ion{S}{2}]
doublet using a pair of Gaussians. The two components were assumed to
have the same velocity width, except at the nucleus where the higher
S/N permitted a fit with fewer constraints.  The line ratios were then
converted to an electron density at each position using the STSDAS
task TEMDEN, for an assumed temperature of $T_e = 10^4$ K.  The
[\ion{O}{1}] \lam6300 line intensity was also measured by fitting a
single Gaussian.  Due to the large difference between the critical
densities of the [\ion{S}{2}] ($\ncrit = 1.6\times10^3$ and
$1.5\times10^4$ \percucm) and [\ion{O}{1}] ($\ncrit = 1.6\times10^6$
\percucm) transitions, the intensity ratio of
[\ion{O}{1}]/[\ion{S}{2}] is sensitive to the nebular density as well.

\begin{figure}
\plotone{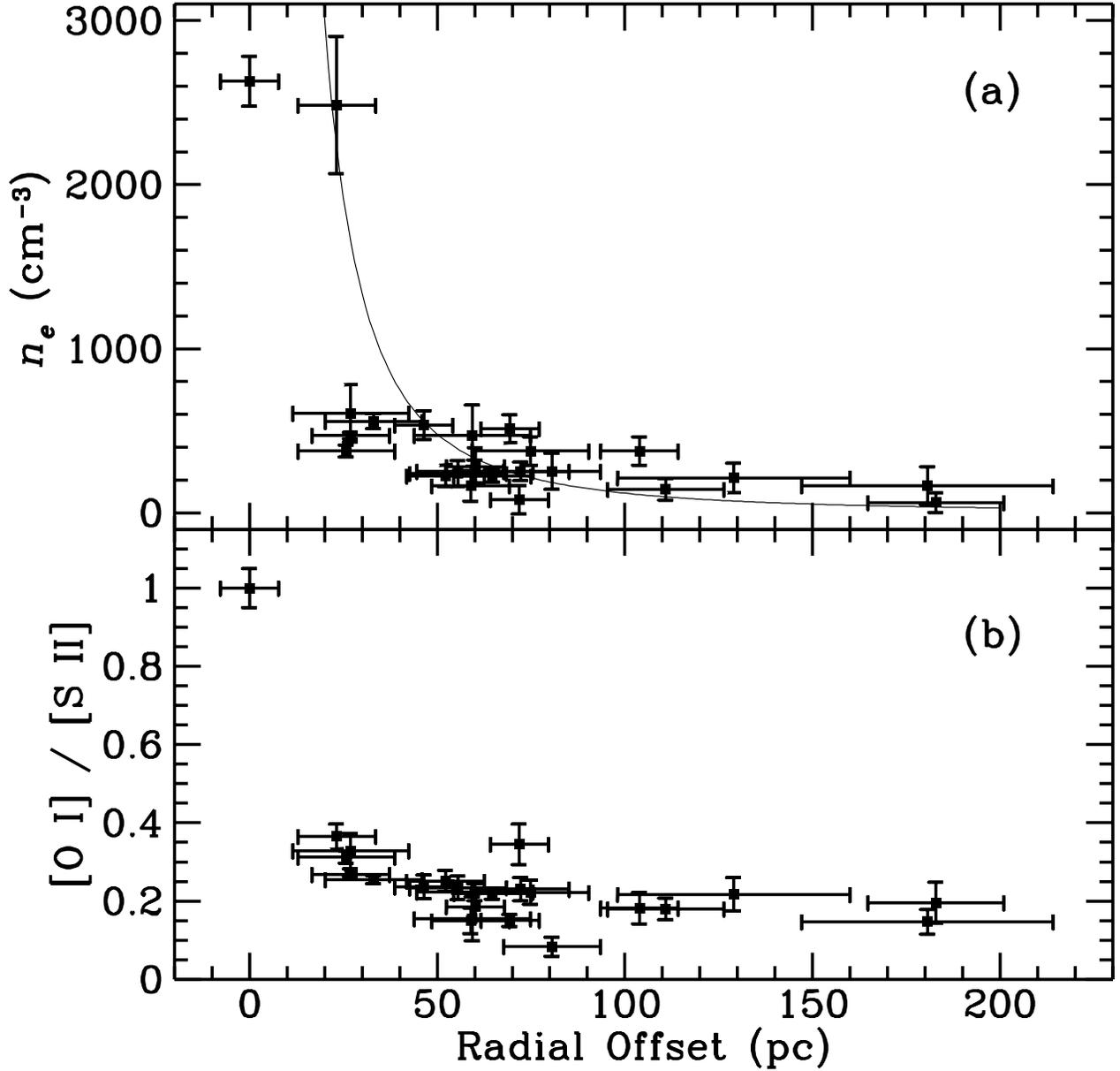}
\caption{Density diagnostics in the NLR of NGC 4579. (\emph{a})
Electron density, determined from the [S II] line ratio, as a function
of position along the slit.  The solid curve represents the model $n_e
\propto r^{-2}$. (\emph{b}) Intensity ratio of [O I] \lam6300 to [S
II] \lam\lam6716, 6731.  In both panels, the horizontal error bars
represent the extraction width used for the emission-line measurement.
Vertical error bars represent the uncertainties in the measured
emission-line intensities.  At the assumed distance of 21 Mpc,
1\arcsec\ corresponds to 101 pc.
\label{fig3}}
\end{figure}

The results of these measurements are shown in Figure \ref{fig3}.  The
electron density measured from the [\ion{S}{2}] lines rises from
$\lesssim200$ \percucm\ in the outer parts of the NLR ($r \gtrsim 100$
pc) to $2630 \pm70$ \percucm\ at the nucleus.  Overall, the trend of
increasing $n_e$ toward the nucleus directly supports the picture of a
radially stratified density in the NLR.  Although the [\ion{S}{2}]
line ratio cannot probe densities greater than $\sim10^4$ \percucm,
the intensity and width of [\ion{O}{1}] at the nucleus implies still
greater densities in the inner NLR.  The [\ion{O}{1}]/[\ion{S}{2}]
ratio exhibits a sharp ``spike'' at the nucleus.  Also, in the
extraction centered on the nucleus, [\ion{O}{1}] \lam6300 (FWHM = 930
\kms) is nearly twice as broad as the [\ion{S}{2}] lines (FWHM = 500
and 570 \kms\ for \lam6716 and \lam6731, respectively), indicating
that the linewidth-\ncrit\ relation continues to hold at the smallest
radii resolvable by \hst.  The difference between the widths of the
two [\ion{S}{2}] lines in the nuclear extraction \citep[an effect first
seen in M81;][]{fs88} also demonstrates the presence of a
linewidth-\ncrit\ correlation within the inner $r=2$ pc.

A simple model for the density structure can be constructed under the
assumption that the ionization parameter $U$ (defined as the ratio of
ionizing photon density to electron density at the illuminated face of
a cloud) is roughly independent of radius \citep{fh84}.  Then, $n_e
\propto r^{-2}$.  The assumption of constant $U$ is certainly an
oversimplification, as the UV spectrum of NGC 4579 demonstrates that
the NLR contains clouds with a range of values of $U$ and $n_e$
\citep{bar96}, and the observed values of $n_e$ show considerable
scatter.  However, the model curve shown in Figure \ref{fig3}
demonstrates that an $r^{-2}$ density law approximates the general
trend of density as a function of radius.  The [\ion{S}{2}] density in
the nuclear extraction falls well below the $r^{-2}$ model prediction,
but as noted already, the [\ion{S}{2}] ratio at the nucleus most
likely underestimates the central density.

To confirm that the [\ion{O}{1}]/[\ion{S}{2}] ratio is indeed
indicative of higher densities at the nucleus, we calculated a grid of
photoionization models using Cloudy \citep{fer98} for a single,
plane-parallel slab of solar-metallicity gas.  Conditions were assumed
to be typical of LINERs and of NGC 4579 in particular: an ionization
parameter of $U = 10^{-3}-10^{-3.5}$ and an ionizing continuum slope
of $\alphaox = 0.9$ \citep{bar96}.  We find that
[\ion{O}{1}]/[\ion{S}{2}] $\geq1$ is only achieved at densities $n_e
\gtrsim 2\times10^4$ \percucm\ for dust-free gas, or at $n_e \gtrsim
10^5$ \percucm\ when gas-phase depletion and dust grains with a
Galactic dust-to-gas mass ratio are included in the calculations.
Thus, the observed [\ion{O}{1}]/[\ion{S}{2}] intensity ratio of
$1.01\pm0.05$ at the nucleus indicates that $n_e$ continues to rise
well above $10^4$ \percucm\ within the unresolved, inner portion of
the NLR, and the inferred density increases further for the
high-velocity material which contributes to the wings of the
[\ion{O}{1}] profile.

\section{Conclusions}

STIS observations have recently revealed very broad double-peaked or
double-shouldered broad \hal\ profiles in three LINERs: NGC 4203
\citep{shi00} and NGC 4450 \citep{ho00} from \hst\ program GO-7361
(which observed a total of 24 targets), and NGC 4579 from program
GO-7403 (with 6 targets observed).  In contrast, no double-peaked
emitters were found during the course of the \citet{hfs97a} Palomar
Observatory survey, which observed 486 nearby galaxies.  Ground-based
spectroscopy is apparently a very inefficient way to detect such
features, except in much more luminous AGNs such as Arp 102B
\citep{ssk83} and Pictor A \citep{he94}.  Among nearby ($D \lesssim20$
Mpc), low-luminosity ($L_{bol} \lesssim10^{43}$ erg s\per) AGNs, NGC
1097 is the only example of a double-peaked emitter discovered from
the ground (Storchi-Bergmann \etal 1993).  It is difficult to
determine meaningful statistics on the incidence of broad
double-peaked emission in the LINER population, but an \hst\
spectroscopic survey of a larger sample of galaxies would provide
useful constraints.

From the width of the very broad \hal\ line in NGC 4579, we infer an
inner radius of $\sim160$ \rs\ for the line-emitting portion of the
accretion disk.  Alternatively, BLR models based on radial outflow
rather than emission from a rotating disk may also be able to explain
the observed \hal\ profile.  Periodic \hst\ monitoring of this galaxy
(and other LINERs having double-peaked emission) would be valuable, as
line-profile variability can provide tests of the accretion-disk and
outflow models.

Finally, the direct detection of a radial density gradient within the
inner NLR of NGC 4579 provides support for the interpretation of
LINERs as photoionized AGNs.

\acknowledgments
 
Support for this work was provided through grant GO-7403 from STScI,
which is operated by AURA, Inc., under NASA contract NAS 5-26555.
NASA grant NAG 3-3556 is also acknowledged.  Research by A.J.B. is
supported by a postdoctoral fellowship from the Harvard-Smithsonian
Center for Astrophysics.  We thank Gary Ferland for making Cloudy
available to the community.  An anonymous referee provided insightful
comments on the manuscript, and we also thank Eliot Quataert for
helpful discussions on accretion disk models.

\end{document}